\documentclass[twocolumn]{aastex631}
\usepackage{lineno}

\newcommand\aastex{AAS\TeX}

\usepackage{xcolor}
\usepackage{wrapfig}
\usepackage{url}
\usepackage{multirow}

\usepackage[T1]{fontenc} 
\usepackage[utf8]{inputenc} 
\usepackage[graphicx]{realboxes}

\usepackage{tipa}

\newcommand{\m}{6.15} % slope of linear fit
\received{\today}
\revised{TBD, 2018}
\accepted{TBD, 2018}

\shorttitle{\aastex\ Rocky Exoplanet and Stellar Compositions}
\shortauthors{Brinkman et al.}

\begin{document}

\title{Revisiting the Relationship Between Rocky Exoplanet and Stellar Compositions: Reduced Evidence for a Super-Mercury Population}

\correspondingauthor{Casey L. Brinkman}
\email{clbrinkm@hawaii.edu}

\author[0000-0002-4480-310X]{Casey L. Brinkman}
\affiliation{Institute for Astronomy, University of Hawai'i, 2680 Woodlawn Drive, Honolulu, HI 96822 USA}

\author[0000-0001-7047-8681]{Alex S. Polanski}
\affiliation{Department of Physics and Astronomy, University of Kansas, Lawrence, KS, USA}

\author[0000-0001-8832-4488]{Daniel Huber}
\affiliation{Institute for Astronomy, University of Hawai'i, 2680 Woodlawn Drive, Honolulu, HI 96822 USA}
\affiliation{Sydney Institute for Astronomy (SIfA), School of Physics, University of Sydney, NSW 2006, Australia}

\author[0000-0002-3725-3058]{Lauren M. Weiss}
\affiliation{Department of Physics and Astronomy, University of Notre Dame, Notre Dame, IN, 46556, USA}

\author[0000-0003-3993-4030]{Diana Valencia}
\affiliation{Centre for Planetary Sciences, University of Toronto, 1265 Military Trail, Toronto, ON, M1C 1A4, Canada}

\author[0000-0002-9479-2744]{Mykhaylo Plotnykov}
\affiliation{Centre for Planetary Sciences, University of Toronto, 1265 Military Trail, Toronto, ON, M1C 1A4, Canada}

\begin{abstract}
Planets and the stars they orbit are born from the same cloud of gas and dust, and the primordial compositions of rocky exoplanets have been assumed to have iron and refractory abundance ratios consistent with their host star. To test this assumption, we modeled the interior iron-to-rock ratio of 20 super-Earth sized (1-1.8R$_{\oplus}$) exoplanets around stars with homogeneously measured stellar parameters. We computed the core mass fraction for each planet and an equivalent ``core mass fraction'' for each host star based on its Fe and Mg abundances. We then fit a linear correlation using two methods (Ordinary Least Squares and Orthogonal Distance Regression) between planetary and stellar core mass fraction, obtaining substantially different slopes between these two methods (m=1.3 $\pm$ 1.0 and m=5.6 $\pm$ 1.6, respectively). Additionally, we find that 75$\%$ of planets have a core mass fraction consistent with their host star to within 1$\sigma$, and do not identify a distinct population of high-density super-Mercuries. Overall, we conclude that current uncertainties in observational data and differences in modeling methods prevent definitive conclusions about the relationship between between planet and host star chemical compositions. 

%While we observe that Venus, Earth, Mars, and smaller bodies such as carbonaceous chondrites have solar abundance ratios of iron to magnesium (Fe/Mg), Mercury is enriched in Fe/Mg while the Moon is depleted, indicating that planet formation processes such as giant impacts and magnetic interaction have the ability to alter primordial rocky compositions.

\end{abstract}


\section{Introduction} 
\label{sec:intro}

Planets are born from the same primordial nebular material as their host star, and it is intuitive to assume, in the absence of reprocessing during formation, that the relative chemical abundances of iron and rock-building elements between star and planet would be similar. There is indeed consistency between solar and planetary abundances in most rocky bodies of the solar system. The relative abundances of most refractory elements (the major elements composing Earth's rocks) reflect that of the Sun \citep{2019Icar..328..287W}. On Venus, the mass ratio of iron to magnesium (Fe/Mg) is not independently constrained, but planetary interior models suggest that Venus likely has a similar bulk composition to that of Earth \citep{1983M&P....29..139Z}. On Mars, the ratio of Fe/Mg equals that of the Sun to within 10-15$\%$ \citep{2003ApJ...591.1220L}. In addition to planets, CI-chondrites---a group of stony meteorites that remain relatively unaltered since the formation of our solar system---have 39 refractory element abundances that are consistent with solar ratios to within 10$\%$ \citep{2003ApJ...591.1220L}, including Fe/Mg to within 2$\%$ \citep{2019AmMin.104..817P}. 

%While this relationship has been challenging to demonstrate for exoplanetary systems, it is well documented in our own solar system. 

However, Mercury differs from the Sun in Fe/Mg by 200-400$\%$ \citep{1980PNAS...77.6973M, Ebel2018}, and therefore underwent additional chemical processing and/or a different formation mechanism from Venus, Earth, and Mars. Meanwhile, the moon is very iron-depleted with only $\sim$3$\%$ of its mass in an iron core \citep{2015LPICo1839.5001S} (compared to $\sim$30$\%$ for the Earth), possibly due to its formation from the debris of a catastrophic giant impact with Earth \citep{2001Natur.412..708C}. The relationship between the elemental abundances of stars and exoplanets can therefore also reveal the dominant pathways of planet formation and potential history of collisions for these worlds. 
%(cite Canup, Ward, Nakajima...)

While some studies have explicitly assumed similar elemental abundance ratios for stars and planets \citep{Dorn2015} others have tried to test it \citep{Plotnykov2020, Adibekyan2021, Schulze2021} albeit with data that was heterogeneous in nature. Even with the plethora of exoplanet discoveries to date, this relationship between planet and stellar compositions has remained difficult to quantify. The first challenge is large uncertainties in density measurements for rocky planets. Rocky planets have small radii and low masses, which are inherently more difficult to measure, and only a handful have masses and radii measured to within 10$\%$ \citep{2017NatAs...1E..56G, 2019ApJ...883...79D, 2020MNRAS.491.2982E, 2021PSJ.....2....1A, 2021Sci...371.1038T, 2021A&A...649A.144S, 2021NatAs...5..775D, Brinkman2023B, 2023A&A...677A..33B}. Previous studies found that the uncertainties---especially in mass---are too large in most cases to draw definitive conclusions about the compositions of individual rocky planet and host star systems \citep{Plotnykov2020, Schulze2021}.\footnote{However, \citet{Schulze2021} did find that most planets and host stars had abundance ratios that agreed to within 1$\sigma$.}

The second challenge is the lack of individual elemental abundance measurements for the host stars. The prevalence of rocky planet discoveries around smaller, cooler host stars, which have spectra dominated by molecular lines, remains a confounding factor in our ability to obtain these measurements, although progress is being made \citep[eg.][]{2023ApJ...949...79H}. In lieu of direct comparison between individual planets and their host stars, \citet{Plotnykov2020} compared the general population of rocky exoplanets to the general population of planet-hosting stars, finding that planets span a wider range in refractory abundances than stars. This result suggests that planet formation pathways that diversify planets from their proto-planetary disk chemistry are necessary (similar to the pathways needed to produce Mercury in our own Solar System). In addition, \citet{Adibekyan2021} (hereafter A+21) performed a one-to-one comparison between small planet elemental abundances and the abundances of their host stars and found a steeper than unity correlation between stellar and planet iron enrichment, suggesting that planet formation processes in iron-rich stars lead to preferentially iron-rich planets.

In this paper, we update the stellar masses and radii for 20 rocky exoplanet host stars based on newly available elemental abundances. We propagate these updated stellar properties to planet masses and radii using literature values of R$_{P}$/R$_{*}$ and radial velocity semi-amplitudes. We compute the core mass fraction (CMF) of each planet and compare these values to the newly determined stellar abundances to investigate the extent to which planet interior compositions correlate with the host star chemical abundance ratios.

\section{Rocky Planet Sample Selection} 
\label{sec:sample}
If a planet hosts a water layer or volatile envelope, it becomes impossible to constrain the relative fraction of iron core to rocky mantle for a planet, even with a precisely measured bulk density \citep{Valencia2007, RogersSeager2010}. A gaseous envelope contributes to the radius of the planet (while contributing minimally to the mass), and prevents us from knowing the mass and radius of the rocky mantle and iron core, giving a degeneracy of possible composition solutions. The masses and radii of small exoplanets suggest that the transition between primarily rocky and gas-enveloped planets occurs at approximately 1.5 $R_{\oplus}$ \citep{2014ApJ...783L...6W, 2015ApJ...801...41R, 2017AJ....154..109F}, with planets smaller than 1.5 $R_{\oplus}$ often having compositions consistent with Earth-like iron-to-silicate ratios \citep{2015ApJ...800..135D}. However, existing super-Earth mass measurements show far more diversity than we observe for rocky planets in our own solar system \citep{2014ApJS..210...20M, 2016ApJ...822...86M, 2019ApJ...883...79D}. This suggests that the interior compositions of Earth and super-Earth sized planets could potentially vary from entirely made of silicate rock, to predominantly made of iron \citep{2019NatAs...3..416B}, with possibly even high mean molecular weight atmospheres \citep{2017AJ....154..232A, 2021ApJ...909L..22K}. 

\begin{figure*}
    \centering
    \includegraphics[width=1.0\textwidth]{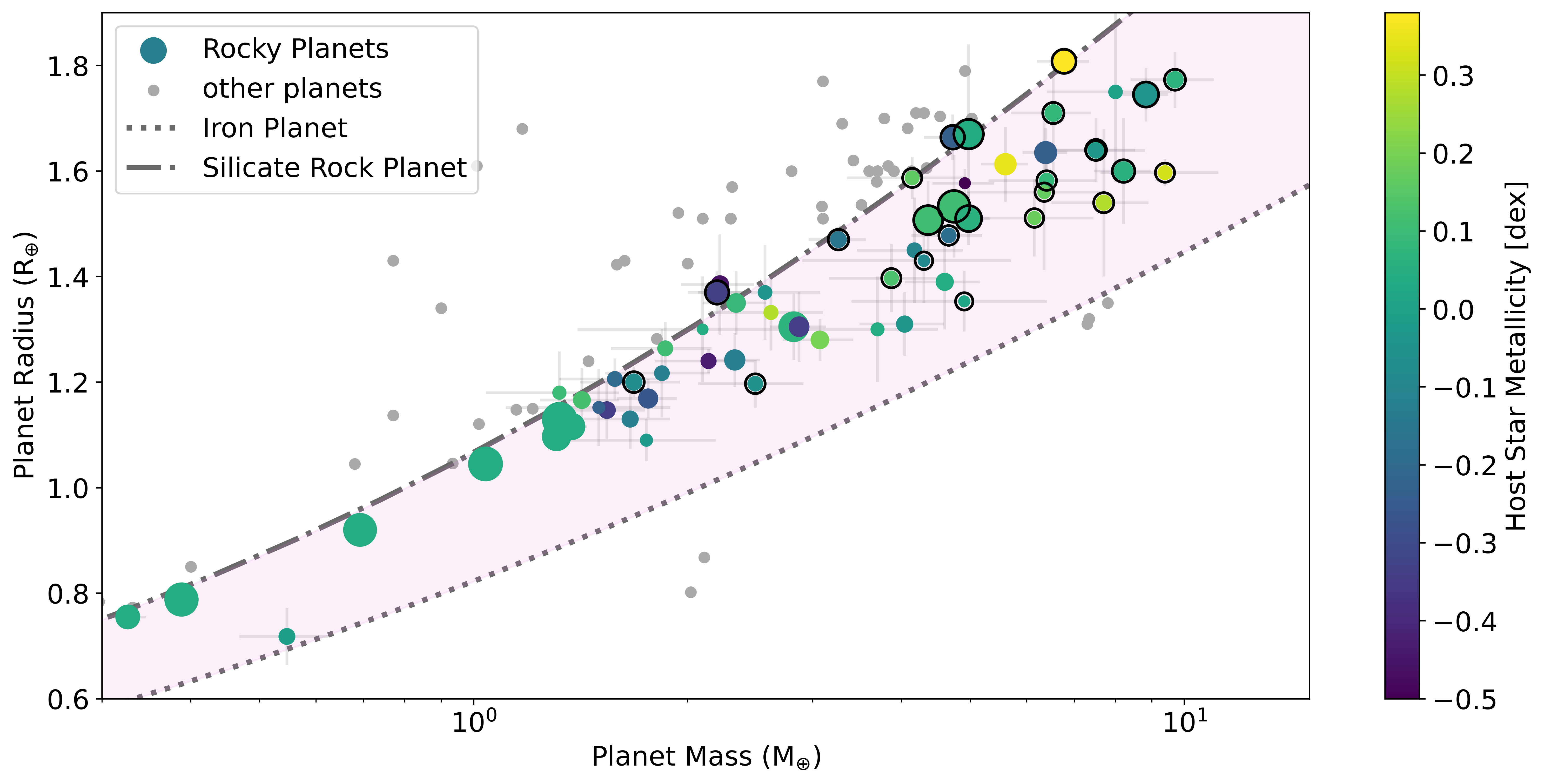}
   
    \caption{Radius vs mass for planets with radii R$<$1.8  R$_{\oplus}$ that have a fractional uncertainty of <30$\%$ in mass and radius, from the NASA Exoplanet Archive (queried 05/06/2023, \citealt{2013PASP..125..989A}). Exoplanets that are consistent with a rocky composition (mixture of iron and rock only, violet shaded region) are shown in color, while those inconsistent with a rocky composition are shown in grey. Planets that are less dense than rock likely have gaseous envelopes, while planets more dense than iron likely have measurement errors, either in mass, radius, or both. Point size scales inversely with the uncertainty in density (larger points have smaller uncertainties), and color indicates host star metallicity. Planets that orbit stars for which we have detailed homogeneously measured abundances for Mg, Si, and Fe are circled in black.}
    \label{fig:massRadius}
\end{figure*}

To curate a list of planets consistent with a rocky composition, we use the NASA Exoplanet Archive (queried 05/06/2023, \citealt{2013PASP..125..989A}). We required planets to have radii R $<$ 1.8 R$_{\oplus}$ and mass and radius measurements with a fractional uncertainty of <30$\%$. Next, we filtered out planets with densities more than 1$\sigma$ outside the bounds for a rocky planet (less dense than silicate rock) using the mass-radius models from \cite{Zeng2019} (Figure \ref{fig:massRadius}). Additionally, we filtered out planets with nonphysically large mass measurements suggestive of compositions more dense than pure iron, and found that these planets primarily had TTV mass measurements instead of RV masses. This resulted in 64 planets, orbiting 51 host stars, as shown in Figure \ref{fig:massRadius}. 

For these planets we performed a careful literature search and selected values for planet properties (most importantly mass and radius) based on the following criteria: (1) self consistent, with mass and radius measurements from the same paper (and using the same stellar mass and radius) where possible, (2) the most up to date and/or inclusive of the most data, and (3) from as few different sources as possible. For example, we used as many mass and radius values as possible from the recent  \cite{2023A&A...677A..33B} catalogue to satisfy multiple of these criteria. 

For 20 of these 64 planets we obtained individual abundance measurements for their host stars. For this sub-sample of planets and host stars we perform a more thorough characterization discussed in Section \ref{sec:abundances}, and do not use literature mass and radius values. For the remainder of this paper we will refer to the full 64 rocky planet sample as the Metallicity Sample, and we will refer to the sub-sample of 20 planets with host star abundances as the \texttt{KeckSpec} sample. Both samples provide useful information: the Metallicity sample is much larger and provides information about planets from a wider range of stellar types, while the \texttt{KeckSpec} sample provides more specific information on compositional similarities between planet and host star.

\section{Updated Host Star and Planet Properties}
\subsection{Stellar Abundances}

\label{sec:abundances}

\begin{figure}
    \centering
    \includegraphics[width=0.46\textwidth]{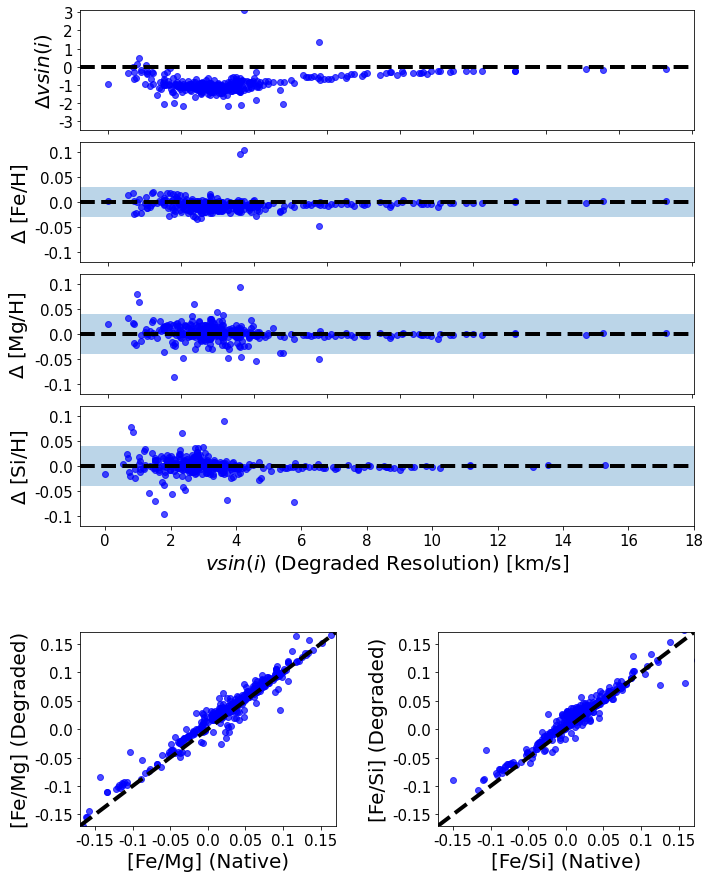}
   
    \caption{The difference in $v\sin{(i)}$, [Fe/H], [Mg/H], and [Si/H] for a sample of spectra that had their spectral resolutions degraded (top four panels) showing that \texttt{KeckSpec} is insensitive to changes in spectral resolution down to a $v\sin{(i)}$ of around 4 km s$^{-1}$. The blue shaded region gives the range of uncertainty quoted for the abundance measurements. The bottom two panels show a comparison of relative abundances between the native spectral resolution and a degraded one.   }
    \label{fig:resolution_test}
\end{figure}
For 19 stars hosting 20 rocky planets we obtained individual elemental abundance measurements, composing our \texttt{KeckSpec} sample. The spectra for these stars were taken on the HIRES instrument at the W. M. Keck Observatory \citep{1994SPIE.2198..362V} without the iodine cell used for deriving radial velocities and all had SNR above 100. Abundances were determined using \texttt{KeckSpec}, a version of the \texttt{Cannon} \citep{2015ApJ...808...16N} that was designed to be applied to iodine-free spectra from the High Resolution Echelle Spectrograph on Keck I \citep{Rice2020}. \texttt{KeckSpec} was trained using a sample of high-quality (SNR > 100) spectra for which $T_{\text{eff}}$, $\log{(g)}$, $v\sin{(i)}$, and abundances for 15 chemical elements were determined \citep{Brewer_2016}. For the sample selected in this work, we used \texttt{SpecMatch Synthetic} \citep{Petigura2015PhDT........82P} to verify that the star's primary stellar parameters ($T_{\text{eff}}$, $\log{(g)}$, $v\sin{(i)}$) and bulk metallicity fall within the parameter space of the \texttt{KeckSpec} training sample. In addition, the retrieved abundances for Mg and Si are all well-within the reliable range of \texttt{KeckSpec} (i.e. the range of abundance values covered by the training sample, Table 3 of \citealt{Rice2020}).

The majority of spectra that comprised the \texttt{KeckSpec} training sample were taken at a resolution of $\sim$70,000, however, there are subsequent iodine-free spectra in our sample that were taken with a wider slit resulting in slightly lower spectral resolution ($\sim$60,000). While this was done in order to enable sufficient sky-subtraction for fainter stars, an examination of possible impacts on our results is warranted. We selected a sub-sample of high-SNR spectra taken at higher spectral resolution and convolved them with a Gaussian kernel to reduce their resolution to $\sim$60,000. The results of this test are shown in Figure \ref{fig:resolution_test}. In general, we find that the residual abundances (native resolution minus degraded) increases with lower measured $v\sin{(i)}$. Changes in the measured abundance value may be expected for lower spectral resolution as the broadening contribution from the stellar rotation decreases. The \texttt{Cannon}'s coefficient vectors for $v\sin{(i)}$ and [X/H] both correlate strongly with the cores of spectral lines (see Figure 6 in \cite{Rice2020}) meaning that when $v\sin{(i)}$ deviates, the abundances also change to compensate. While the increased scatter in abundances is small, typically half of the quoted uncertainty on these measurements, on a per-spectrum basis the degraded resolution affects the elements of interest to this study (Fe, Mg, Si) in similar ways. The result is that relative abundances, e.g. [Fe/Si], remain comparable between native and degraded spectral resolutions. There appears to be a potential offset in [Fe/H] abundances between native and degraded resolutions, so to test the effects of this potential offset on our stellar ``CMF" values we purturbed all [Fe/H] values by 0.02 dex and found no significant change in our results. Further analysis of these systems and a KeckSpec analysis of $>$5000 additional stars can be found in Polanski et al. \textit{in prep}.

While \texttt{KeckSpec} provides measurements for 15 elements, the abundance measurements we primarily consider for planet composition are magnesium, iron, and silicon, and for the analysis of host star populations we consider $\alpha$ abundances. Magnesium is widely accepted as the best rock building element to compare with primordial compositions, since it is less volatile than silicon and far less volatile than oxygen \citep{Yakovlev2018}.

%\begin{longtable*}{|c|c|c|c|c|c|c|c|c|c|c|}\hline
\begin{table*}
\centering

\caption{Stellar Parameters}
\label{table:stellar}

\begin{tabular}{|c|c|c|h|c|c|c|c|c|C|c|c|c|c|} \hline 
 Star Name & T$_{eff}$    & Log(g)  &   Luminosity    & Radius          & Mass            & [Fe/H]  & [Mg/H] & [Si/H]  &[$\alpha$/H ] & Fe/Mg         & Fe/Si         & Equivalent \\  
           & [K]         &          & [L$_{\odot}$]   & [R$_{\odot}$]   & [M$_{\odot}$]   & [dex]   & [dex]  & [dex]   &  [dex]       &             &               & CMF \\ \hline
HD 136352  & 5730      & 4.5        & 0.95 $\pm$ 0.04 & 0.98 $\pm$ 0.04 & 0.86 $\pm$ 0.05 & -0.31   &-0.20   & -0.23   &  0.11        & 1.5 $\pm$ 0.2 & 1.5 $\pm$ 0.3 & 0.24 \\ 
HD 15337   & 5131      & 4.4        & 0.45 $\pm$ 0.01 & 0.85 $\pm$ 0.03 & 0.85 $\pm$ 0.04 & 0.12    & 0.13   & 0.12    & -0.00        & 1.9 $\pm$ 0.2 & 1.7 $\pm$ 0.7 & 0.30 \\ 
HD 219134  & 4858      & 4.6        & 0.28 $\pm$ 0.04 & 0.76 $\pm$ 0.04 & 0.78 $\pm$ 0.03 & 0.08    & 0.09   & 0.11    &  0.02        & 1.9 $\pm$ 0.2 & 1.6 $\pm$ 0.7 & 0.30 \\ 
HD 3167    & 5254      & 4.4        & 0.53 $\pm$ 0.02 & 0.87 $\pm$ 0.03 & 0.85 $\pm$ 0.04 & 0.03    & 0.06   & 0.05    &  0.02        & 1.8 $\pm$ 0.2 & 1.7 $\pm$ 0.6 & 0.29 \\ 
K2-106     & 5479      & 4.4        & 0.80 $\pm$ 0.03 & 0.98 $\pm$ 0.04 & 0.91 $\pm$ 0.05 & 0.05    & 0.04   & 0.03    &  -0.01       & 2.0 $\pm$ 0.2 & 1.8 $\pm$ 0.6 & 0.31 \\
K2-229     & 5196      & 4.6        & 0.39 $\pm$ 0.02 & 0.79 $\pm$ 0.02 & 0.83 $\pm$ 0.04 & 0.02    & 0.01   & 0.02    &  0.01        & 1.9 $\pm$ 0.3 & 1.7 $\pm$ 0.6 & 0.30 \\
K2-265     & 5395      & 4.4        & 0.65 $\pm$ 0.02 & 0.92 $\pm$ 0.04 & 0.89 $\pm$ 0.05 & 0.04    & 0.05   & 0.04    &  0.00        & 1.9 $\pm$ 0.2 & 1.7 $\pm$ 0.6 & 0.30 \\ 
K2-291     & 5503      & 4.4        & 0.66 $\pm$ 0.02 & 0.90 $\pm$ 0.03 & 0.93 $\pm$ 0.04 & 0.08    & 0.03   & 0.06    &  -0.02       & 2.2 $\pm$ 0.3 & 1.8 $\pm$ 0.6 & 0.33 \\
K2-36      & 4841      & 4.6        & 0.25 $\pm$ 0.01 & 0.72 $\pm$ 0.02 & 0.75 $\pm$ 0.03 & -0.03   &-0.06   & 0.03    &  0.02        & 2.0 $\pm$ 0.3 & 1.5 $\pm$ 0.5 & 0.32 \\ 
K2-38      & 5678      & 4.3        & 1.26 $\pm$ 0.05 & 1.15 $\pm$ 0.05 & 1.05 $\pm$ 0.05 & 0.23    & 0.22   & 0.24    &  -0.01       & 1.9 $\pm$ 0.3 & 1.7 $\pm$ 0.9 & 0.30 \\
Kepler-10  & 5671      & 4.4        & 1.06 $\pm$ 0.11 & 1.06 $\pm$ 0.06 & 0.89 $\pm$ 0.05 & -0.17   &-0.07   & -0.09   &  0.09        & 1.5 $\pm$ 0.2 & 1.5 $\pm$ 0.4 & 0.25 \\
Kepler-20  & 5463      & 4.4        & 0.77 $\pm$ 0.19 & 0.97 $\pm$ 0.12 & 0.89 $\pm$ 0.04 &  0.01   & 0.03   & 0.03    &  0.01        & 1.8 $\pm$ 0.2 & 1.7 $\pm$ 0.6 & 0.29 \\
Kepler-21  & 6174      & 4.3        & 4.87 $\pm$ 0.04 & 1.92 $\pm$ 0.09 & 1.33 $\pm$ 0.05 &  0.00   &-0.08   & 0.02    &  0.00        & 2.2 $\pm$ 0.3 & 1.9 $\pm$ 0.6 & 0.35 \\
Kepler-406 & 5607      & 4.3        & 1.04 $\pm$ 0.04 & 1.07 $\pm$ 0.04 & 1.01 $\pm$ 0.04 & 0.21    & 0.21   & 0.22    &   0.00      & 1.9 $\pm$ 0.3 & 1.7 $\pm$ 0.9 & 0.30 \\
%Kepler-65  & 6177      & 4.2        & 0.31 $\pm$ 0.01 & 0.75 $\pm$ 0.02 & 0.78 $\pm$ 0.03 & 0.20    & 0.12   & 0.17    & 2.3 $\pm$ 0.3 & 1.9 $\pm$ 0.9 & 0.35 \\
Kepler-78  & 5003      & 4.5        & 0.31 $\pm$ 0.01 & 0.75 $\pm$ 0.02 & 0.78 $\pm$ 0.03 & -0.02   &-0.11   & -0.01   & -0.01        & 2.3 $\pm$ 0.3 & 1.7 $\pm$ 0.5 & 0.35 \\
Kepler-93  & 5642      & 4.5        & 0.80 $\pm$ 0.03 & 0.93 $\pm$ 0.04 & 0.89 $\pm$ 0.05 & -0.16   & -0.1   & -0.13   &  0.04        & 1.8 $\pm$ 0.2 & 1.6 $\pm$ 0.4 & 0.28 \\ 
%Kepler-99  & 4774      & 4.5        & 0.95        & 0.98        & 0.86        & 0.24    & 0.19   & 0.24    & 2.1 $\pm$ 0.3 & 1.7 $\pm$ 1.0 & 0.33 \\
TOI-1444   & 5377      & 4.4        & 0.63 $\pm$ 0.02 & 0.91 $\pm$ 0.03 & 0.88 $\pm$ 0.05 & 0.04    & 0.06   & 0.04    &  0.01        & 1.8 $\pm$ 0.2 & 1.7 $\pm$ 0.6 & 0.29 \\
TOI-561    & 5357      & 4.6        & 0.54 $\pm$ 0.02 & 0.83 $\pm$ 0.02 & 0.76 $\pm$ 0.04 & -0.40   &-0.22   & -0.19   &  0.18        & 1.3 $\pm$ 0.2 & 1.1 $\pm$ 0.2 & 0.21 \\
Wasp-47    & 5482      & 4.3        & 1.05 $\pm$ 0.05 & 1.12 $\pm$ 0.05 & 1.01 $\pm$ 0.05 & 0.36    & 0.34   & 0.30    &  -0.01       & 2.0 $\pm$ 0.3 & 2.0 $\pm$ 1.3 & 0.31 \\ \hline

\end{tabular}
%\begin{tablenotes}
      \small
      \\
      Abundances, effective temperature, and surface gravity were measured from high-resolution spectroscopy. Luminosities, radii, and masses were derived using \texttt{isoclassify}. We give the mass ratios of Fe/Mg and Fe/Si in each star, calculated in section \ref{sec:abundances}, along with the stellar equivalent of ``Core Mass Fraction.'' Uncertainties: $\sigma_{T_{eff}}$=100 K, $\sigma_{Log(g)}$=0.1, $\sigma_{[Fe/H]}$=0.04 dex, $\sigma_{[Mg/H]}$=0.05 dex, $\sigma_{[Si/H]}$=0.04 dex, $\sigma_{[\alpha/H]}$=0.07 dex, $\sigma_{CMF}$=0.03. For reference, the Solar values for element mass ratios are Fe/Mg=1.9, Fe/Si=1.77, and CMF=0.30. 
%\end{tablenotes}

%\end{longtable*}
\end{table*}

\begin{figure*}
    \centering
    \includegraphics[width=1.0\textwidth]{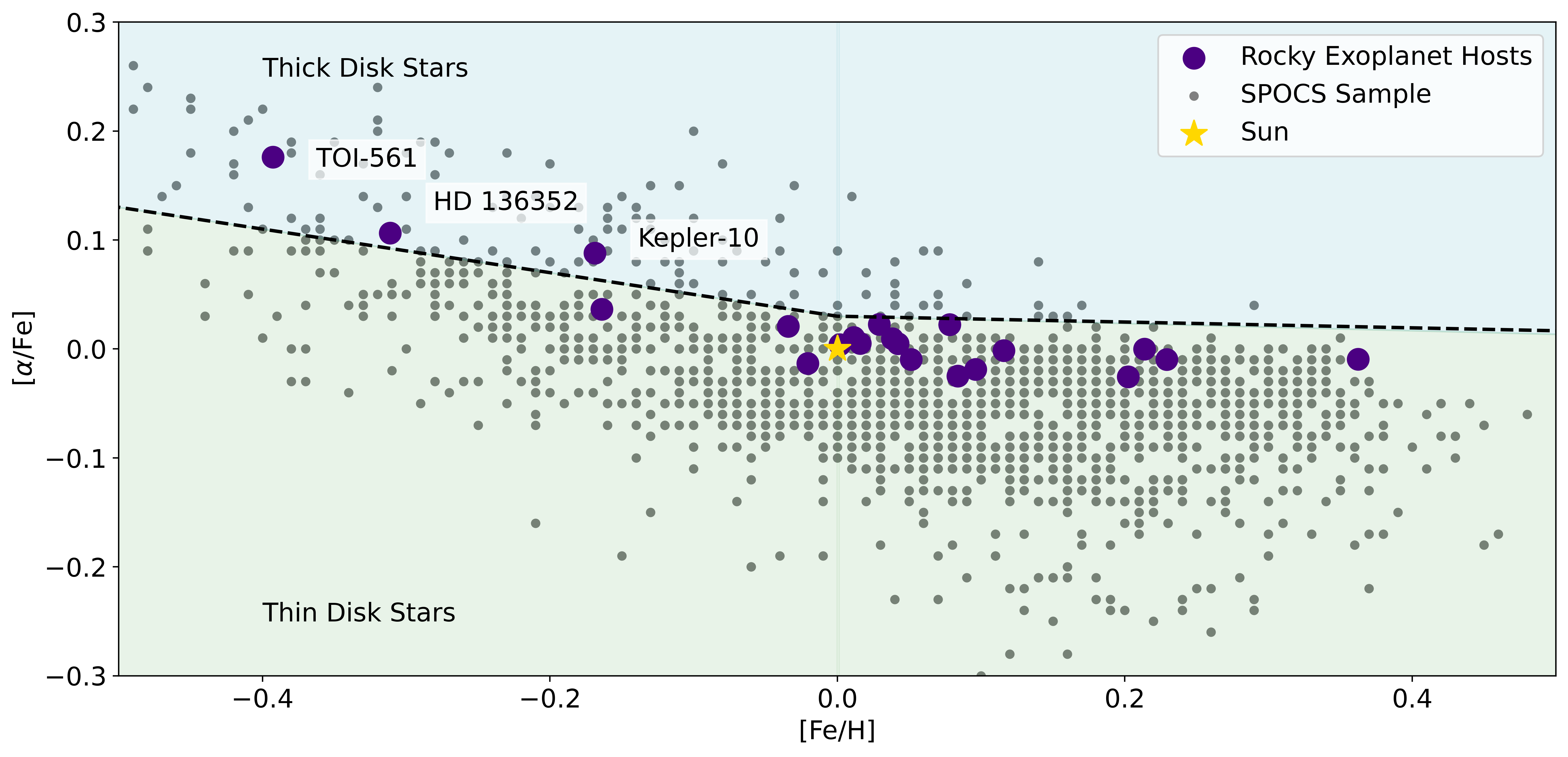}
   
    \caption{Alpha abundances [$\alpha$/Fe] as a function of [Fe/H] for each host star in our \texttt{KeckSpec} sample (indigo), along with a larger sample of exoplanets from the SPOCS catalogue \citep{Brewer_2018}. The black dashed line approximately separates the galactic thin disc and thick disc populations \citep{2021AJ....161...56W}. The three stars in our sample that fall above this line and were likely born in the galactic thick disc are labeled. }
    \label{fig:alpha}
\end{figure*}

With these abundance measurements we can also analyze our rocky planet host stars in terms of galactic population. We use $\alpha$ abundances (computed as the average of Mg, Si, Ti, and Ca abundances) and compare [$\alpha$/Fe] and [Fe/H] for each star in our sample as shown in Figure \ref{fig:alpha}. We find that three stars in our sample (TOI-561, Kepler-10, and HD 136352) have higher $\alpha$ abundances relative to iron that are characteristic of the chemically defined thick disc \citep{2006MNRAS.367.1329R}, in agreement with previous findings \citep{2014ApJ...789..154D, 2020AJ....160..129K, 2021AJ....161...56W}. 

\subsection{Masses and Radii}
\label{sec:MandR}
%The determination of stellar masses and radii depends sensitively on stellar metallicity, and the new elemental abundances we obtained for 19 planet-hosting stars (\texttt{KeckSpec} sample) provided an opportunity to update their stellar masses and radii. This will allow us to update the planet masses and radii as well, ensuring that all of the stellar parameters used in our both planet and host star measurements are homogeneously determined and self-consistent. 

To update the stellar masses and radii given our new abundance measurements, we used the ``direct mode'' of \texttt{isoclassify} \citep{2017ApJ...844..102H} to calculate the luminosity of each star using T$_{\mathrm{eff}}$, [M/H], and log($g$) from \texttt{KeckSpec}, along with the Gaia DR3 parallax, 2MASS K-band magnitude, a 3D dust map and bolometric corrections. Typically, \texttt{isoclassify} uses stellar evolution models that use solar-scaled $\alpha$ abundances, implicitly assuming that [Fe/H]=[M/H]. However, given the number of non-solar abundance measurements we obtained, we use [$\alpha$/Fe] and [Fe/H] from \texttt{KeckSpec} to compute overall [M/H] using the calibration from \cite{1993ApJ...414..580S}:
\begin{equation}
    \mathrm{[M/H]} = \mathrm{[Fe/H]} + \mathrm{log_{10}}(0.694 × 10^{\mathrm{[\alpha/Fe]}} + 0.306)
\end{equation}
We then used these derived luminosities, T$_{\mathrm{eff}}$, and [M/H] in the ``grid mode'' of \texttt{isoclassify} to infer the mass and radius of each star from a grid of MIST isochrones \citep{2016ApJ...823..102C}. Our newly derived luminosities, masses and radii of the planet-hosting stars are listed in Table \ref{table:stellar}. 

We then re-calculated planet mass (m) using our homogeneously derived stellar masses (M$_{*}$) along with literature values for radial velocity semi-amplitude (K), orbital period (P), inclination (i), and eccentricity (e) using the relation:
\begin{equation}
    K=\frac{28.4329}{\sqrt{1-e^{2}}}\frac{m\mathrm{sin}(i)}{M_{\mathrm{Jup}}}(\frac{m+M_{*}}{M_{\odot}})^{-2/3}(\frac{P}{1 \mathrm{yr}})^{-1/2}.
\end{equation} 
We also calculated planet radius using our homogeneously derived stellar radii and literature values for R$_{P}$/R$_{*}$. The updated planet masses and radii for the \texttt{KeckSpec} sample are presented in Table \ref{table:planetparams}, along with the adopted literature values for K and R$_{P}$/R$_{*}$.

\section{Chemical Abundance Ratios of Planets and Stars}
\subsection{Planet Core Mass Fraction}
\label{sec:CMF}
Using the mass and radius of each planet we calculated the core mass fraction (CMF), a measure of the mass fraction of the iron core to the total planet mass \citep[e.g.][]{2007ApJ...669.1279S, 2014ApJ...787..173H, 2016ApJ...819..127Z}. For reference, the Earth is $32.5\%$ iron by mass, giving it a CMF of 0.325. We used \texttt{SuperEarth} \citep{2006Icar..181..545V, Plotnykov2020} to model the interior composition of each planet. The package solves equations of state for iron and various rock-building minerals to match the mass and radius values provided. To first order, the code assumes the planet has two primary, differentiated layers (an iron core and a rocky mantle). \texttt{SuperEarth} then refines this approximation by assuming a four-layer mantle composition (upper mantle, transition zone, lower mantle, and lower-most mantle) like that of the Earth, distinguished by the mineral phase boundaries determined by the pressure and temperature of the mantle. As a further refinement, {\tt SuperEarth} takes user-provided molar fractions of silica inclusion in the iron core (here assumed to be 0), as well as iron molar fractions in the mantle (here assumed to be 0.1). To test the significance of these assumptions, we varied the silica inclusion and iron molar fractions by 10$\%$ and found no significant change in our results. 

%An outcome of modeling the planet interiors with \texttt{SuperEarth} is that we can readily compute Fe/Mg and Fe/Si mass ratios for each planet because \texttt{SuperEarth} uses a fixed mineral composition in each mantle layer. We calculate Fe/Mg by taking Fe/Si from \texttt{SuperEarth} and assuming a ratio of Mg/Si=0.97361 \citep{Plotnykov2020}. 

We calculated CMF for both the full Metallicity sample of rocky planets (as shown in Figure \ref{fig:massRadius}), as well as for our \texttt{KeckSpec} sub-sample. We used the best literature values for planet mass and radius (described above) to calculate CMF for the Metallicity sample, and for planets in our \texttt{KeckSpec} sub-sample we used the masses and radii we calculated from homogeneous host star measurements (Table \ref{table:planetparams}). We assumed that these planets are rocky and contain only iron and rock for the purpose of this paper, but cannot ensure they have negligible water layer or gaseous envelope. Therefore, the CMF we computed for each planet is a minimum CMF value, as the presence of a low-density layer would require a larger iron core to produce the same planet density we observe. 

To compute CMF uncertainties, we drew 1000 values for the mass and radius of each planet from Gaussian distributions centered on best-fit values with 1$\sigma$ error bars. We then propagated these values through \texttt{SuperEarth} and took the standard deviation of the resulting CMF distribution, (after ensuring that distribution is also Gaussian). We assumed no correlation between planet mass and radius, although there likely is one because the stellar mass and radius associated with the best-fit model isochrones are correlated, and so our Monte Carlo draws represent conservative uncertainty estimates.

%\begin{longtable*}{|c|c|c|c|c|c|c|c|c|c|c|}\hline
\begin{table*}
\begin{centering}
\caption{Planet Parameters}
\begin{tabular}{|c|c|c|c|c|c|h|h|} \hline
 Planet Name &  R$_{p}$/R$_{*}$          & K                     & Radius            & Mass             &  CMF              & Fe/Mg           & Fe/Si        \\  
             &                           & [m/s]                 & [R$_{\oplus}$]    & [M$_{\oplus}$]   &                   &                 &                                   \\  \hline
HD 136352 b  & 0.0144 $\pm$ 0.0003$^{LD}$& 1.46 $\pm$ 0.12$^{LD}$& 1.66 $\pm$ 0.04  & 4.7 $\pm$ 0.4     & 0.25 $\pm$ 0.18   & 0.02 $\pm$ 0.50  & 0.02 $\pm$ 0.46  \\  
HD 15337 b   & 0.0176 $\pm$ 0.0006$^{G}$ & 3.08 $\pm$ 0.44$^{G}$ & 1.64  $\pm$ 0.06 & 6.5 $\pm$ 1.2     & 0.22  $\pm$ 0.17  & 1.88 $\pm$ 2.00 & 1.94 $\pm$ 1.96  \\  
%HD 213885 b  & 1.745 $\pm$ 0.0516 & 8.83$\pm$ 0.66   &  0.36 $\pm$ 0.12  & 2.24 $\pm$ 1.20 & 2.30 $\pm$ 1.18 & \cite{2020MNRAS.491.2982E} \\  
HD 219134 b  & 0.018 $\pm$ 0.001$^{S}$   & 2.38 $\pm$ 0.08$^{S}$ & 1.60 $\pm$ 0.06  & 4.7  $\pm$ 0.2    &  0.35 $\pm$ 0.28  & 0.63 $\pm$ 0.71 & 0.65 $\pm$ 0.72  \\ 
HD 219134 c  & 0.0177 $\pm$ 0.0009$^{S}$ & 1.70 $\pm$ 0.08$^{S}$ & 1.51 $\pm$ 0.05  & 4.4 $\pm$ 0.2     &  0.17 $\pm$ 0.27  & 1.43 $\pm $0.88 & 1.47 $\pm$ 0.88  \\ 
HD 3167 b    & 0.0171 $\pm$ 0.0009$^{VB}$& 3.56 $\pm$ 0.15$^{B}$ & 1.67  $\pm$ 0.17 & 5.0 $\pm$ 0.2     &  0.03 $\pm$ 0.09  & 0.23 $\pm$ 1.10 & 0.36 $\pm$ 1.10  \\ 
K2-106 b     & 0.0157 $\pm$ 0.0007$^{A}$ & 6.5  $\pm$ 0.5$^{B}$  & 1.73 $\pm$ 0.04  & 8.2  $\pm$ 0.8    &  0.41 $\pm$ 0.21  & 2.07 $\pm$ 1.08 & 2.13 $\pm$ 1.08  \\ 
%K2-141 b     & 1.510 $\pm$ 0.05   & 4.97  $\pm$ 0.35 &  0.45 $\pm$ 0.11  & 3.17 $\pm$ 1.52 & 3.25 $\pm$ 1.52 & \cite{2023A&A...677A..33B} \\ 
K2-229 b     & 0.0140 $\pm$ 0.0002$^{D}$ & 2.2 $\pm$ 0.36$^{D}$  & 1.20 $\pm$ 0.05  & 2.5  $\pm$ 0.4    &  0.56 $\pm$ 0.13  & 8.29 $\pm$ 182.77& 8.51 $\pm$ 182.77  \\ 
K2-265 b     & 0.0160 $\pm$ 0.0004$^{L}$ & 3.34 $\pm$ 0.43$^{L}$ & 1.71  $\pm$ 0.11 & 6.5  $\pm$ 0.8    &  0.39 $\pm$ 0.17  & 0.93 $\pm$ 2.97 & 0.95 $\pm$ 2.97   \\ 
K2-291 b     & 0.0160 $\pm$ 0.0003$^{D}$ & 3.31 $\pm$ 0.56$^{D}$ & 1.58 $\pm$ 0.04  & 6.4   $\pm$ 1.1   &  0.45 $\pm$ 0.14  & 3.10 $\pm$ 2.10 & 3.19 $\pm$ 2.10   \\ 
K2-36 b      & 0.0183 $\pm$ 0.0007$^{MD}$& 2.85 $\pm$ 0.92$^{B}$ & 1.43  $\pm$ 0.08 & 3.9   $\pm$ 1.1   &  0.38 $\pm$ 0.21  & 2.18 $\pm$ 19. 2& 2.23 $\pm$19.22   \\
K2-38 b      & 0.0133 $\pm$ 0.0007$^{T}$ & 3.02 $\pm$ 0.43$^{B}$ & 1.66 $\pm$ 0.10  & 7.7  $\pm$ 1.2    &  0.41  $\pm$ 0.24 & 35.54 $\pm$ 232.36& 36.51 $\pm$ 232.36  \\ 
Kepler-10 b  &0.01268 $\pm$ 0.00004$^{D}$& 2.34 $\pm$ 0.21$^{B}$ & 1.47  $\pm$ 0.03 & 3.3  $\pm$ 0.3    &  0.06 $\pm$ 0.25  & 0.63 $\pm$ 0.73 & 0.65 $\pm$ 0.73   \\ 
%Kepler-107 c & 1.597 $\pm$ 0.026  & 9.39  $\pm$ 1.77 &  0.69  $\pm$ 0.15 & 8.20 $\pm$24.89 & 8.42 $\pm$ 24.89 & \cite{2019NatAs...3..416B} \\ 
Kepler-20 b  & 0.0177 $\pm$ 0.0005$^{LB}$& 4.23 $\pm$ 0.54$^{B}$ & 1.77 $\pm$ 0.05  & 9.7  $\pm$ 1.3    &  0.14  $\pm$ 0.43 & 1.06 $\pm$ 1.18 & 1.09 $\pm$ 1.18   \\ 
Kepler-21 b  &0.00789$\pm$ 0.00005$^{LM}$& 2.7 $\pm$ 0.5$^{B}$   & 1.64 $\pm$  0.02 & 7.5  $\pm$ 1.3    &  0.39  $\pm$ 0.17 & 3.08 $\pm$ 1.81 & 3.17 $\pm$ 1.80   \\ 
Kepler-406 b & 0.0125 $\pm$ 0.0003$^{M}$ & 2.4 $\pm$ 0.8$^{W}$   & 1.56 $\pm$ 0.15  & 6.4 $\pm$ 1.4     &  0.52 $\pm$ 0.18  & 3.47 $\pm$ 1672.32 & 3.56 $\pm$ 1672.32 \\ 
%Kepler-65 d  &                  &       & 1.587 $\pm$ 0.04   & 4.14  $\pm$ 0.79 &  -0.01 $\pm$ 0.22 & 0.21 $\pm$ 0.85 & 0.21 $\pm$ 0.85  \\ 
Kepler-78 b  & 0.0151 $\pm$ 0.0002$^{D}$ & 1.8 $\pm$ 0.3$^{B}$   & 1.20  $\pm$ 0.03 & 1.7  $\pm$ 0.3    &  0.01  $\pm$ 0.17 & 1.64 $\pm$ 1.08 & 1.69 $\pm$ 1.08  \\ 
Kepler-93 b  &0.01475$\pm$0.00006$^{CD}$ & 1.9 $\pm$ 0.2$^{B}$   & 1.48 $\pm$ 0.02  & 4.7  $\pm$ 0.5    &  0.34  $\pm$ 0.15 & 1.46 $\pm$ 1.18 & 1.50 $\pm$ 1.18  \\ 
%Kepler-99 b  &                  &       & 1.511 $\pm$ 0.073  & 6.15  $\pm$ 1.3  &  0.56  $\pm$ 0.25 & 4.69 $\pm$67.03 & 4.82 $\pm$ 67.03 \\ 
TOI-1444 b   & 0.0141 $\pm$ 0.0006$^{D}$ & 3.3 $\pm$ 0.6$^{D}$   & 1.40 $\pm$ 0.06  & 3.9  $\pm$ 0.7    &  0.39  $\pm$ 0.21  & 2.93 $\pm$ 8.72 & 3.01 $\pm$ 8.72 \\ 
TOI-561 b    & 0.0151 $\pm$ 0.0004$^{CB}$& 2.2 $\pm$ 0.2$^{CB}$  & 1.37  $\pm$ 0.04 & 2.2   $\pm$ 0.2   &  -0.10  $\pm$ 0.17 & -0.06$\pm$ 0.51 & -0.06 $\pm$ 0.51 \\ 
Wasp-47 e    & 0.0146 $\pm$ 0.0001$^{BB}$& 4.6 $\pm$ 0.4$^{BB}$  & 1.81 $\pm$ 0.03  & 6.8  $\pm$ 0.6    &  -0.01  $\pm$ 0.18 & 0.12 $\pm$ 0.37 & 0.12 $\pm$ 0.37 \\ \hline

\end{tabular}
%\begin{tablenotes}
      \small 
      \\
       Mass and Radius values are given for each planet consistent with a rocky composition (selection process described in section \ref{sec:sample}) for which we also have measured host star refractory abundances. Mass and Radius values were calculated using the literature R$_{p}$/R$_{*}$ and RV semi-amplitude along with homogeneously measured host star masses and radii (Table \ref{table:stellar}). We use these values to compute the Core Mass Fraction (CMF) and mass ratio of Fe/Mg and Fe/Si using {\tt SuperEarth} \citep{2006Icar..181..545V, Plotnykov2020}. Literature Reference Key: LD=\cite{2021NatAs...5..775D}, G=\cite{2019ApJ...876L..24G}, S=\cite{2021AJ....161..117S}, VB=\cite{2022A&A...668A..31B}, B=\cite{2023A&A...677A..33B}, A=\cite{2021PSJ.....2..152A}, D=\cite{2019ApJ...883...79D}, L=\cite{2018A&A...620A..77L}, MD=\cite{2019A&A...624A..38D}, T=\cite{2020A&A...641A..92T}, LB=\cite{2016AJ....152..160B}, LM=\cite{2016AJ....152..204L}, M=\cite{2016ApJ...822...86M}, W=\cite{2024ApJS..270....8W}, CD=\cite{2015ApJ...800..135D}, CB=\cite{Brinkman2023B}. BB=\cite{2022AJ....163..197B}
   % \end{tablenotes}
\label{table:planetparams}
\end{centering}

%\end{longtable*}
\end{table*}

\subsection{Stellar Equivalent ``CMF''}
\label{sec:Star_CMF}
To compare compositions between star and planet, we must express the stellar abundances and planet compositions in equivalent quantities. We computed the mass ratio of Fe/Mg from these abundance measurements, and then computed the stellar equivalent value of planet CMF. Stellar abundance measurements are given in the form:
\begin{equation}
  % [E/H]=$log_{10}(($n$(Fe)/$n$(H))_{*}/($n$(Fe)/$n$(H))_{\odot}$
  [E/H]=\log_{10}(\frac{n(E)/n(H)_{*}}{n(E)/n(H)_{\odot}})
\end{equation}
where n(E)/n(H)$_{*}$ is the number density of an element E relative to hydrogen relative to the Sun. To turn this into an absolute number density for the star (not relative to the Sun), we used the number density for each element in the Sun relative to hydrogen, given as:
\begin{equation}
    A(E)=12 + \log_{10}(n(E)/n(H)_{\odot})
\end{equation}
Using A(Fe)=7.46 $\pm$ 0.04, A(Mg)=7.55 $\pm$ 0.03, and A(Si)=7.51 $\pm$ 0.03 \citep{2021A&A...653A.141A}, we solved for the number density of these three elements relative to hydrogen (n(E)/n(H)$_{\odot}$). We then used these values to solve for n(E)/n(H)$_{*}$ in Equation 3. 

With values for the number density of each element, we then calculated the mass of each of these elements relative to Hydrogen using the atomic weights of each species (55.85 u for Fe, 24.3 u for Mg, and 28 u for Si). We then found the mass ratios Fe/Mg and Fe/Si in each host star (given in Table \ref{table:stellar}). 

We then translated this Fe/Mg ratio in the host star into an equivalent value for ``Core Mass Fraction'', where the core here is not the core of the star but an expression of iron mass to total iron and rock-building element mass (assuming the same ratio of Mg/Si$\approx$0.9 as the planet mineral composition and that of Earth, \citealt{1995ChGeo.120..223M}). This is done through a function in \texttt{SuperEarth} that uses the inverse process of computing Fe/Mg ratio from CMF for the planets as described in Section \ref{sec:sample}. This allows us to compare equivalent quantities for star and planet composition. These equivalent CMFs are listed in Table \ref{table:stellar}, along with uncertainties which were calculated by drawing 1000 values of [E/H] from Gaussian distributions centered on our measurements with 1$\sigma$ widths, and then computing the Fe/Mg mass ratio and equivalent CMF for each draw and taking the standard deviation of the resulting CMF distributions for each star. Due to the fact that the uncertainties on [Fe/H], [Mg/H], and [Si/H] were consistent across stars, the CMF Uncertainty on each star is consistent as well. 

\section{Planet and Host Star Correlations}
\label{sec:Correlations}
\subsection{Metallicity}
 \begin{figure*}
    \centering
    \includegraphics[width=1.0\textwidth]{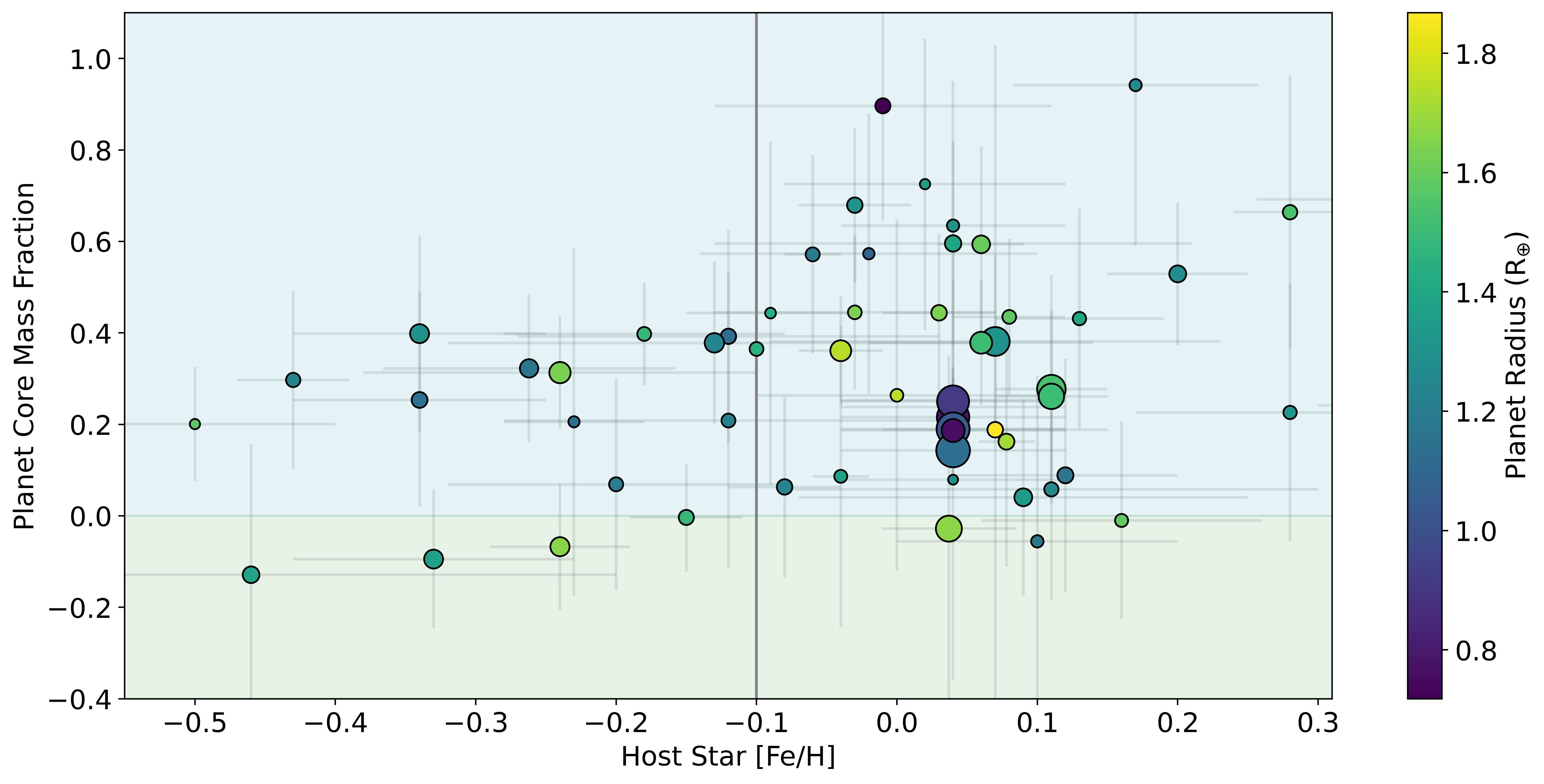}
   
    \caption{Planet Core Mass Fraction (CMF) of the planet as a function of host star metallicity. The color corresponds to planet radius. CMF values below 0 (green shaded region) mean that the planet density is too low to be composed entirely of rock and iron, and likely has a low-density component such as a gaseous envelope or water layer. The vertical grey line divides host stars below and above the median metallicity in the solar neighborhood ([Fe/H]=-0.1 dex).}
    \label{fig:metallicity}
\end{figure*}

Roughly a third of our Metallicity sample orbit M-type host stars. Due to their cool atmospheres, the optical spectra of M-dwarfs are typically dominated by complex molecular lines (such as TiO) that make measuring individual abundances challenging \citep{2022ApJ...927..123S}. While most M-type stars do not have full detailed abundance measurements, [Fe/H] metallicity can be measured through calibration from available lines (such as Na I and Ca I) using low resolution spectroscopy (e.g. as \cite{2012ApJ...748...93R}), or direct measurements of Fe I lines with high-resolution near-IR spectra (e.g. as \cite{2012A&A...542A..33O}). Individual $\alpha$ and refractory abundances can be measured in some M-dwarfs through calibration from synthetic spectra \citep[e.g..][]{2017ApJ...851...26V, 2023ApJ...944...41I}, and high-resolution Near-IR spectra \citep{2022ApJ...927..123S}. 

Refractory abundances are the most direct way to compare planet and stellar compositions, however metallicity does give a general idea about how iron rich the proto-planetary environment must have been. Figure \ref{fig:metallicity} shows planet core mass fraction as a function of host star metallicity for our full Metallicity sample. We observe that there are fewer planets around metal poor host stars than metal rich stars, and in particular we observe few metal-rich high density planets around metal-poor host stars. There are no rocky planets with a CMF above 0.5 around stars below a metallicity of -0.1, which coincides with the median value for stars in the solar neighborhood \citep{2016MNRAS.455..987C}. Meanwhile, above -0.1 dex planets have CMF ranging from $\approx$0--1.0, suggesting that metal-rich host stars can form a wider diversity of planets. 

We perform a Kolmogorov-Smirnov test to examine the statistical significance of the dearth of iron-rich planets around iron-poor stars. We divide our Metallicity sample into two populations: those orbiting stars with metallicity $>$-0.1, and those orbiting stars with metallicity $<$-0.1. We compare the cumulative distributions of their CMFs, finding a p-value of 0.036. This provides tentative evidence that iron-rich planets form less readily around iron-poor stars, regardless of the specific ratio of iron to rock-building elements within the host star.

\subsection{Iron vs. Rock-Building Abundances}
Figure 4 shows the CMF of planets versus the equivalent CMF of the host stars for our \texttt{KeckSpec} sample with homogeneously measured stellar parameters. We observe a wider spread in planet CMF (0-0.6) than host star CMF (0.21-0.35) in agreement with \citep{Plotnykov2020}, suggesting that planet formation mechanisms likely diversify a planet's primordial composition. 

\begin{figure*}

    \includegraphics[width=1.0\textwidth]{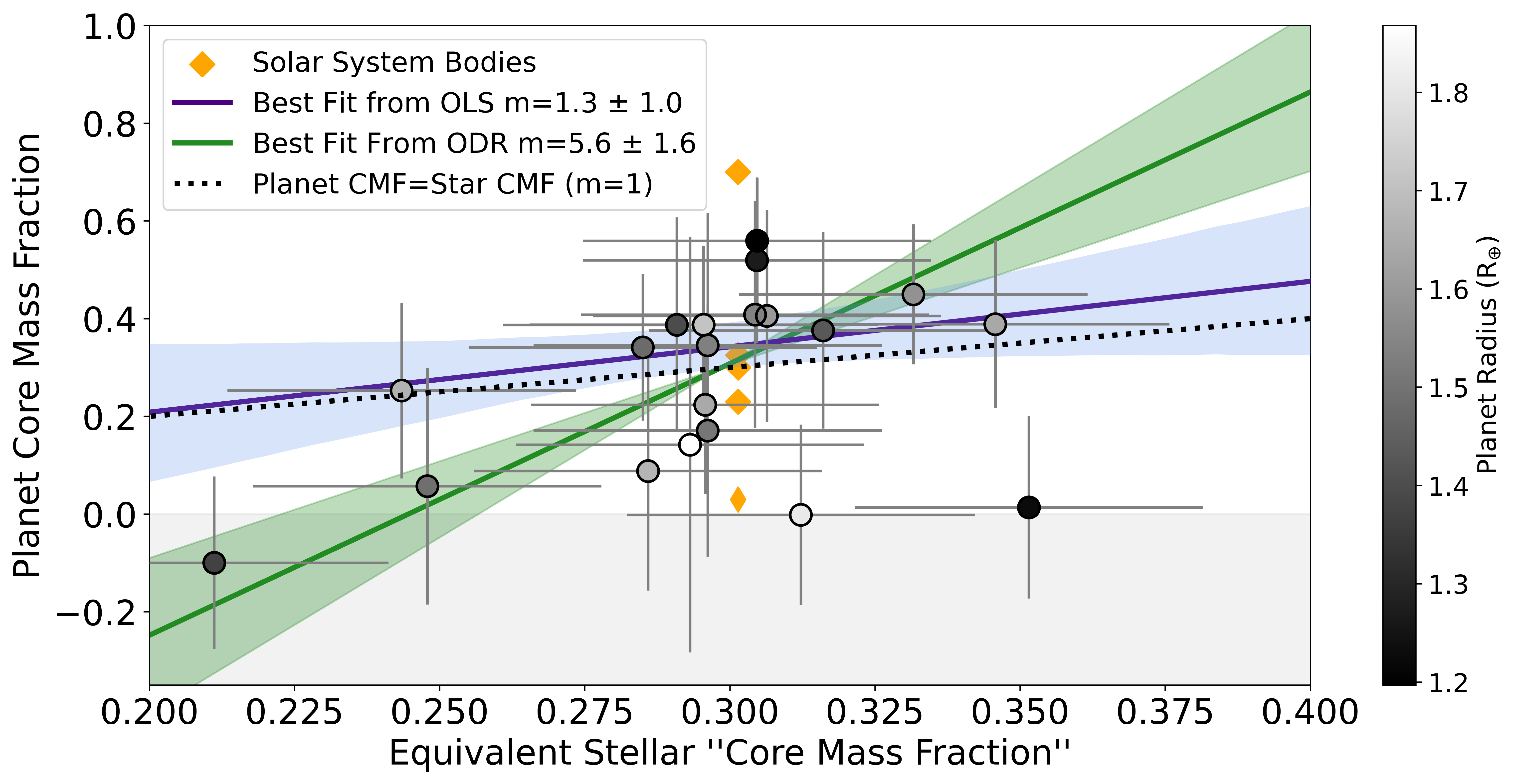}

    \caption{Core Mass Fraction (CMF) of rocky exoplanets vs equivalent CMF (see section \ref{sec:Star_CMF}) for their host stars, shaded by planet radius. The CMF for five solar system bodies are shown at the equivalent CMF of the Sun (from top to bottom: Mercury, Earth, Venus, Mars, and the Moon). The dotted black line shows the one-to-one correspondence of CMF in stars and planets, which is where planets would fall if they inherited the exact Fe/Mg ratio from their host star. Our linear best fit models are shown in indigo (Ordinary Least Squares, OLS) and green (Orthogonal Distance Regression, ODR) with 1$\sigma$ confidence intervals on each.}
    \label{fig:FeMg}
\end{figure*}

We know that planetary bulk compositions have been diversified from their primordial composition in at least one system: the solar system, (yellow diamonds in Figure \ref{fig:FeMg}). While Earth, Venus, and Mars all have similar CMFs, Mercury is significantly enriched in iron (CMF$\approx$0.7) while the moon is significantly depleted (CMF$\approx$0.03). The diversity in CMF that we observe for Solar System bodies is comparable to the spread that we see amongst planets orbiting stars with similar equivalent stellar CMF as the sun. It seems unlikely that the solar system is the only place where planet bulk compositions have been diversified, and the greater spread in CMF of our planets than the stars they orbit is likely evidence of such diversification processes happening in other planetary systems.

While the spread in planet CMF is quite large, Figure \ref{fig:FeMg} shows that planet CMF does appear to correlate slightly with stellar CMF. To quantify a possible relationship, we perform a linear fit (form $y=mx+b$) where $y$ is planet CMF, $x$ is host star equivalent CMF, $m$ is the slope, and $b$ is the intercept. We use the \texttt{curve\_fit} ordinary least squares (OLS) functionality in {\tt SciPy} to find the best-fit slope and intercept. We then use a Monte Carlo approach to estimate the uncertainties on these parameters by drawing values for CMF for each planet and star from Gaussian distributions and repeating the fit. As part of this process, we reject all values for planet density that produce a negative CMF (grey shaded region of Figure \ref{fig:FeMg}). This allows us to include information about planets that have a high probability of being rocky (such as TOI-561 b) without biasing our fit toward unphysically low CMFs. We repeat this 1000 times and recover a median slope of $m=1.3$ $\pm$ 1.0 and intercept of $b=-0.05$ $\pm$ 0.28. This slope is consistent with a slope of one (stars and planets have equal iron-to-silicates ratios) and only 1.3 $\sigma$ larger than a slope of zero (all planets have the same iron-to-silicate ratio, regardless of the composition of their host star). This suggests only weakly that planets and their host stars have similar compositions.  

To test the sensitivity of our results to the fitting methodology, we perform an independent analysis using orthogonal distance regression (ODR) with {\tt SciPy.odr}. ODR has the additional benefit of accounting for uncertainty on data in both $x$ and $y$ without utilizing the bootstrapping technique we applied to our OLS fit. We use the same values for planet CMF and host star equivalent CMF as in our OLS fit, and use the uncertainties listed in Tables \ref{table:stellar} and \ref{table:planetparams}. The choice of ODR instead of OLS produces a slope of $m=5.6$ $\pm$ 1.6, and an intercept of $b=-1.4$ $\pm$ 0.5. This slope is >3$\sigma$ larger than 0, suggesting a more statistically significant result than the slope we obtained with the OLS method. The ODR-obtained slope is also >2$\sigma$ larger than a slope of 1, indicating a steep relationship between planet and star compositions (discussed more in section \ref{sec:discussion}). The OLS and ODR linear fits are shown in Figure \ref{fig:FeMg}. 

We then tested the sensitivity of our results on our choice of Fe/Mg vs Fe/Si. We used \texttt{SuperEarth} to compute the stellar equivalent CMF using Fe/Si instead of Fe/Mg, then we computed the linear relationship between planet and host star CMF using these alternative stellar CMF values. We obtained results that are consistent with with what we get using Fe/Mg derived stellar equivalent CMFs using OLS (m=1.5 $\pm$ 1.0) and ODR (m=7.1 $\pm$ 1.9). This shows that our result is robust to the choice of using Mg or Si as the tracer for rock-building abundances.

In the \texttt{KeckSpec} sample, there is a dearth of iron-rich planets orbiting stars with low Fe/Mg, consistent with our findings for the full Metallicity sample. There are only three planets orbiting stars with equivalent stellar CMF $<$ 0.275, but all of these have a CMF $<$0.3. These three planets (TOI-561 b, Kepler-10, and HD 136352) orbit stars identified as thick-disc stars based on their stellar chemical abundances and kinematics \citep{2021AJ....161...56W, 2020AJ....160..129K, 2011ApJ...729...27B} as shown in Figure \ref{fig:alpha}. With such a small sample we cannot infer much about the diversity of rocky planet compositions around metal-poor hosts (in particular thick-disc hosts), but the discovery of more such systems will help us understand rocky planet formation in these environments.

\section{Discussion}
\label{sec:discussion}
\subsection{The Star-Planet Relationship (or lack thereof)}
\subsubsection{Dependence on Fitting Method}
An important result from our study is that the relationship between planet and stellar iron fraction depends sensitively on the fitting algorithm used, which directly affects the interpretation of the star-planet relationship. A slope of unity (m=1) suggests that planets tend to have the same compositions as their host star, while a slope steeper than unity indicates that iron-rich stars not only host more iron-rich planets, but planets enriched far beyond their nebular composition. A steep slope suggests that iron-enriching processes (discussed below) happen disproportionately around metal-rich stars, which \cite{Adibekyan2021} suggest could be due to the fact that iron-rich stars form planets at a higher rate than metal-poor stars, leading to increased likelihood of planet collisions and therefore an increased likelihood of iron-enriched planets. Additionally, a slope steeper than unity suggests that metal-poor stars more often host planets depleted in iron well beyond their primordial compositions. 

Our OLS fit yielded only weak evidence that planet and host star compositions are correlated, whereas the ODR fit suggests that the slope relating planet iron enrichment to stellar iron enrichment is steeper than unity and statistically significant. Previously, \cite{Adibekyan2021} used an ODR fit and found a slope of m=6.3 $\pm$ 1.2, similar to our work, concluding that planet iron enrichment is steeper than what could be expected from stellar composition alone. However, the differences between the OLS and ODR fits in our study are concerning, and we do not find compelling examples of iron-rich planets orbiting iron-rich stars and iron-poor planets orbiting iron-poor stars (discussed below).

%We test which uncertainties (planet or stellar CMF) contribute more significantly to the ODR slope. To do this, we set the uncertainties on planet CMF to zero (keeping stellar CMF uncertainties as listed in Table \ref{table:stellar}) and recover a slope of m=4.9 $\pm$ 1.0. We then replace the uncertainties on stellar CMF with zero (keeping planet CMF uncertainties as listed in Table \ref{table:planetparams}) and recover a slope of m=2.5 $\pm$ 1.3. We find it curious that without uncertainties in both host star or planet CMF, the slope is shallower than with uncertainties on both. 

To test the likelihood of ODR finding a steep and significant slope from the data by chance, we created a sample of 20 simulated planet and host star CMFs randomly drawn from uniform distributions. We bounded the distributions by the smallest and largest CMF in our \texttt{KeckSpec} sample for both planets and host stars, and then fit a linear model using ODR to the simulated data. We repeated this process 1000 times, and found a median slope of m=1.1 with a standard deviation of 3. The slope we measured for our \texttt{KeckSpec} sample is 1.5 $\sigma$ larger than the median of the randomly generated samples, giving a $\sim$13$\%$ chance of measuring the slope we observe by chance.
%However, we find a median uncertainty on the slope in each trial of only 1.2, and 
Additionally, we found that in 6$\%$ of trials the slope measured by ODR is >3$\sigma$ greater than 0. We conclude that there is 6-13$\%$ probability for ODR to find a steep and statistically significant slope by chance given the sample and its uncertainties. %where there is no underlying correlation (or potentially a much shallower correlation in the case where planets and host stars have the same composition and m=1). 

To further investigate the strength of the linear correlation between planet and host star CMF we calculate the pearson correlation coefficient using the corrcoef function in \texttt{NumPy}. We find a correlation coefficient of 0.40, which suggests a weak to moderate correlation. This is consistent with the fact that OLS fit is within 2$\sigma$ of 0 and not a statistically significant result.

\subsubsection{Low CMF Planets}

The majority of planets in our sample (75$\%$) have CMF values consistent with their host star to within 1$\sigma$. We identify 5 planets with a CMF that differs from that of its host star by $>$1$\sigma$: 3 planets with a lower CMF than their host star (Kepler-78 b, TOI-561 b, and Wasp-47 e), and 2 planets with a higher CMF (K2-229 b and Kepler-406 b). 

In our Solar System we do not find any significantly iron depleted planets, but for the Moon a low CMF is likely the result of giant impacts \citep{1976LPI.....7..120C, 2014AREPS..42..551A}. For exoplanets, lower than expected CMF has been suggested as a result of (1) mantle stripping due to giant impacts during formation (more efficient for low-mass planets, \citealt{2020MNRAS.493.4910S, 2022ApJ...940..144S}) (2) sufficiently hot proto-planetary disks (where high-temperature refractory material forms planets depleted of iron \citealt{Dorn2019}), or (3) gaseous envelopes (potentially dominated by high mean molecular weight species, \citealt{2017AJ....154..232A, 2021ApJ...909L..22K}). If hot enough (T$_{eq}>$2000 K), a significant melt fraction in the rocky mantle can also lower the density of a planet \citep{Bower2019}, but this requires at least an atmosphere if not a gaseous envelope to sufficiently redistribute heat across the planet surface. 

%HD 3167 b has a CMF of 0.03, and previous studies have suggested a composition that is not entirely rocky, but could instead host either water or magma oceans along with a water steam envelope \citep{2022A&A...668A..31B}.

Three of our outliers have low CMFs, and two even have ``Negative'' CMFs, which indicates that the planet is too low density to be fully rocky and likely hosts a gaseous envelope. To explain the low density of TOI-561 b, WASP-47 e, and Kepler-78 b, a gaseous envelope of high mean molecular weight species (and potentially large mantle melt fraction) has been suggested \citep{Brinkman2023B, 2022MNRAS.511.4551L, 2023arXiv230808687P, 2022AJ....163..197B}. It is possible that all of three planets are also iron-depleted through mantle stripping or irradiated proto-planetary disks (processes (1) and (2) above). These three planets span a wide range in stellar equivalent CMF (0.2-0.35), and are orbiting both the most iron rich (Kepler-78) and iron poor (TOI-561) host stars, not just around metal-poor stars as would be suggested by a steep slope such as our ODR fit. 

%due to the high equilibrium temperature ($\approx$2500 K) and small planet radius (1.37 R$_{\oplus}$). The mass and radius we adopt for WASP-47 e also produce a planet with a negative CMF and therefore likely hosts a low-density component (such as a steam envelope as suggested in \citealt{2022AJ....163..197B}).

While we have selected planets that are consistent within 1$\sigma$ of having a rocky composition (as discussed in Section \ref{sec:sample}), there is the potential for more than these three planets in our sample to have low-density components and therefore bias our results. To study this possibility, in Figure \ref{fig:FeMg} we set the greyscale of each point proportional to planet radius. R$_{P}$=1.5 R$_{\oplus}$ has been empirically suggested as the transition point between purely rocky planets and those that host H/He envelopes \citep{2014ApJ...783L...6W, 2017AJ....154..109F}, although this boundary is not definitive (e.g. TOI-561 b has R$_{P}$<1.5 R$_{\oplus}$) and planets larger than 1.5 R$_{\oplus}$ are often consistent with a purely rocky composition. If planets with envelopes are present in our sample, the low density of the envelope would result in an interpretation of a low (and possibly negative) CMF. However, the planets with CMF furthest from that of their host star in our sample mostly have radii below R$_{P}$=1.5 R$_{\oplus}$, at both the lowest and highest density values. Although we cannot rule out the possible bias of unintentionally including some small planets with gaseous envelopes, the choice to include planets with radii up to 1.8 R$_{\oplus}$ does not appear to bias our results. 

\subsubsection{High CMF Planets}
Rocky exoplanets with enriched iron cores are often called Super-Mercuries \citep{2010ApJ...712L..73M}. For Mercury, iron enrichment could be the result of giant impacts \citep{2004ApJ...613L.157A}, electromagnetic disk processing \citep{2021PEPS....8...39M}, or a combination of factors \citep{Ebel2018}. Giant impacts have also been suggested as a method of enriching exoplanets in iron \citep{2010ApJ...712L..73M, 2012ApJ...745...79L}, along with a series of smaller impacts \citep{2018ApJ...865...35C, 2019ApJ...881..117S}, mantle evaporation \citep{2018NatAs...2..393S}, temperature-dependent iron enrichment in the planet-forming disk \citep{1972E&PSL..15..286L}, and planet-star \citep{2017MNRAS.465..149J} or planet-planet \citep{2020ApJ...888L...1D} tidal interactions. Single impact events are now considered highly unlikely to produce Mercury's core \citep{2019AJ....157..208C, 2022MNRAS.515.5576F} and a combination of phenomena are likely needed to explain not only Mercury, but high CMF exoplanets across the galaxy.

\begin{table}[]
\begin{flushleft}

\caption{Super-Mercury Comparison}
\label{table:mercuries}
\begin{tabular}{|l|l|l|l|}
\hline
 Planet Name                  &              & A21         & This Work      \\\hline
 \multirow{4}{*}{K2-38 b}     & Mass         & 7.31 $\pm$ 1.4  & 7.7 $\pm$ 1.2  \\\cline{2-4}
                              & Radius       & 1.54 $\pm$ 0.14 & 1.66 $\pm$ 0.10  \\\cline{2-4}
                              & CMF          & 0.63 $\pm$ 0.3   & 0.4 $\pm$ 0.2    \\\cline{2-4}
                              & Stellar CMF  & 0.31 $\pm$ 0.03 & 0.30 $\pm$ 0.03 \\\hline
\multirow{4}{*}{K2-106 b}     & Mass         & 8.36 $\pm$ 0.95 & 8.21 $\pm$ 0.75  \\\cline{2-4}
                              & Radius       & 1.52 $\pm$ 0.16 & 1.73 $\pm$ 0.04  \\\cline{2-4}
                              & CMF          & 0.75 $\pm$ 0.3  & 0.31  $\pm$ 0.2   \\\cline{2-4}
                              & Stellar CMF  & 0.32 $\pm$ 0.03 & 0.37 $\pm$ 0.03  \\\hline
\multirow{4}{*}{Kepler-406 b} & Mass         & 6.4 $\pm$ 1.4   & 6.4 $\pm$ 1.4 \\\cline{2-4}
                              & Radius       & 1.44 $\pm$ 0.03 & 1.56 $\pm$ 0.15 \\\cline{2-4}
                              & CMF          & 0.74 $\pm$ 0.2  & 0.52 $\pm$ 0.18 \\\cline{2-4}
                              & Stellar CMF  & 0.31 $\pm$ 0.03 & 0.30 $\pm$ 0.03 \\\hline
\multirow{4}{*}{K2-229 b}     & Mass         & 2.59 $\pm$ 0.43 & 2.49 $\pm$ 0.42 \\\cline{2-4}
                              & Radius       & 1.16 $\pm$ 0.05 & 1.20 $\pm$ 0.05\\\cline{2-4}
                              & CMF          & 0.69 $\pm$ 0.2  & 0.56 $\pm$ 0.13 \\\cline{2-4}
                              & Stellar CMF  & 0.31 $\pm$ 0.03 & 0.30 $\pm$ 0.03 \\\hline

\end{tabular}
%\begin{tablenotes}
      \small
      Planet mass and radius values for four planets identified as Super-Mercuries in \cite{Adibekyan2021} (A21). We show the mass and radius as published in A21 along with those measured in this work (cross listed with Table \ref{table:planetparams}). All masses are listed in M$_{\oplus}$ and radii in R$_{\oplus}$. Both planet and stellar CMF listed here were calculated using our methodology (discussed in sections \ref{sec:CMF} and \ref{sec:Star_CMF}) and not those published in A21. 
   % \end{tablenotes}
\end{flushleft}

\end{table}

Our \texttt{KeckSpec} sub-sample does not appear to have a distinct population of super-Mercuries, separated from the other small planets with a gap. This is in contrast to the finding of \cite{Adibekyan2021}, who identified a group of five super-Mercuries (K2-38 b, K2-106 b, K2-229 b, Kepler-107 c, Kepler-406 b) that were distinct from the rest of the sample. They suggested that this cannot be due to giant impacts alone because collisional stripping should produce a continuous distribution of CMFs. Four of these five planets are in our \texttt{KeckSpec} sample, but we observe no such gap in planet CMF in Figure \ref{fig:FeMg}, and we do not find that these planets are distinctly higher-density than the rest of the super-Earths. 

Our measurements for mass and radius suggest that K2-229 b and Kepler-406 b have a larger CMF than their host stars by 2$\sigma$ and 1.2$\sigma$, respectively, and could therefore be enriched in iron relative to their primordial compositions; however, they do not appear to be outliers from the rest of the super-Earths. We find that K2-38 b and K2-106 b have CMFs that are consistent with that of their host star to within 1 $\sigma$. Kepler-107 c was not included in our sample because we did not have a sufficiently high SNR spectrum to measure host star abundances of non-iron elements. We find no planets with CMFs that are >2$\sigma$ larger than their host star. Therefore, we do not find compelling evidence of a population of iron-enriched super-Mercuries or a population that is distinct from the population super-Earths. 

\begin{figure*}
\centering
\gridline{\fig{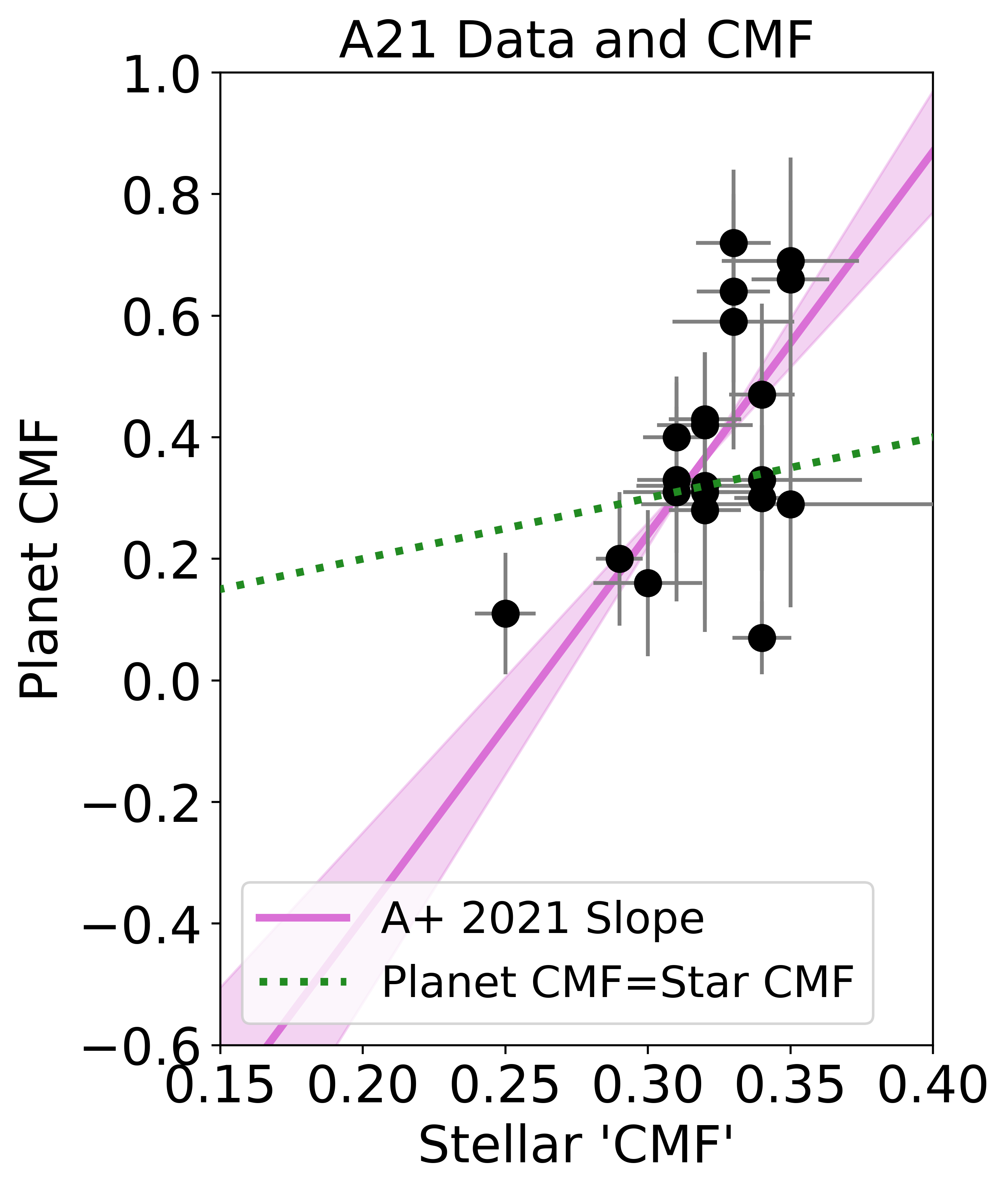}{0.31\textwidth}{}
          \fig{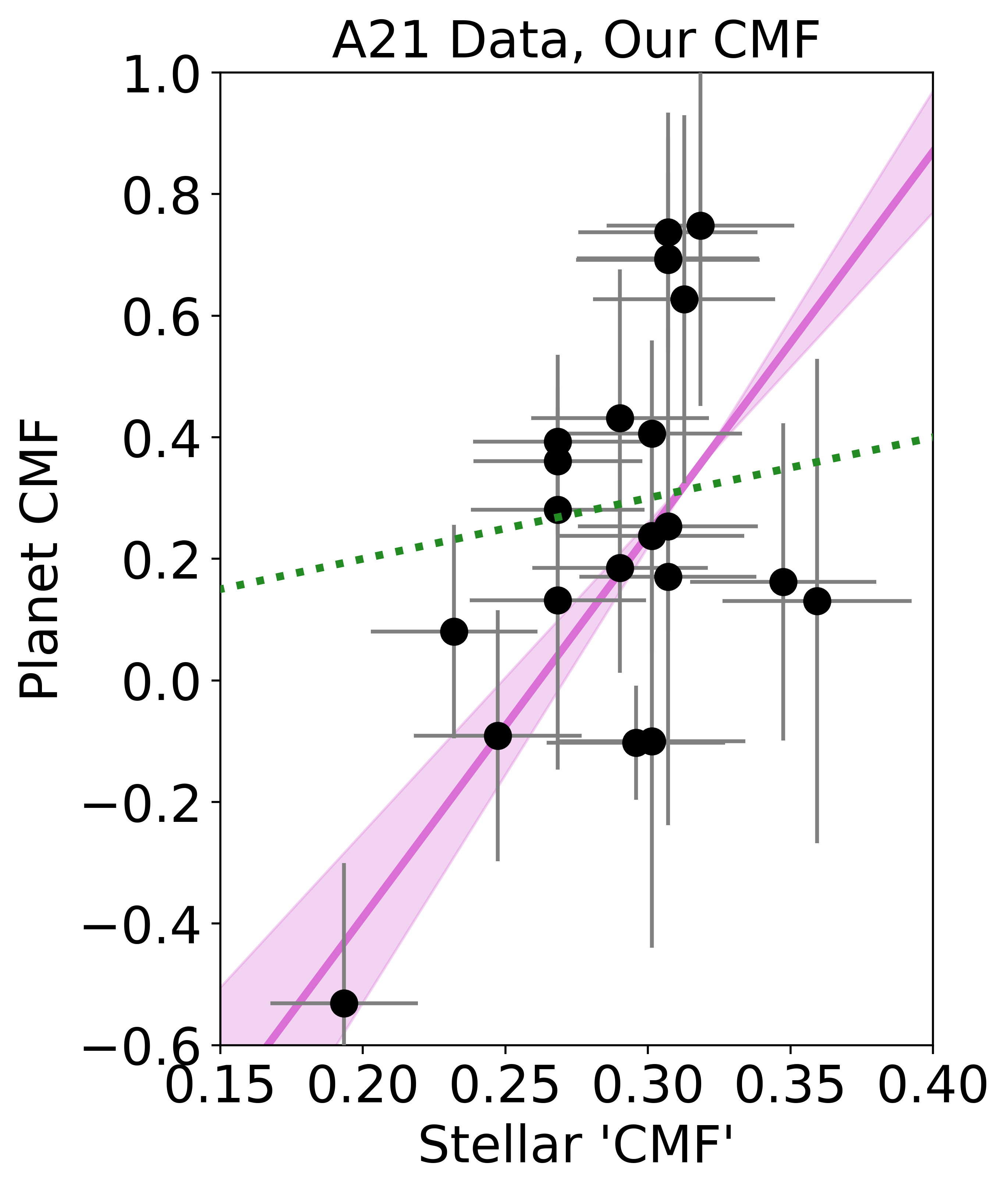}{0.31\textwidth}{}
          \fig{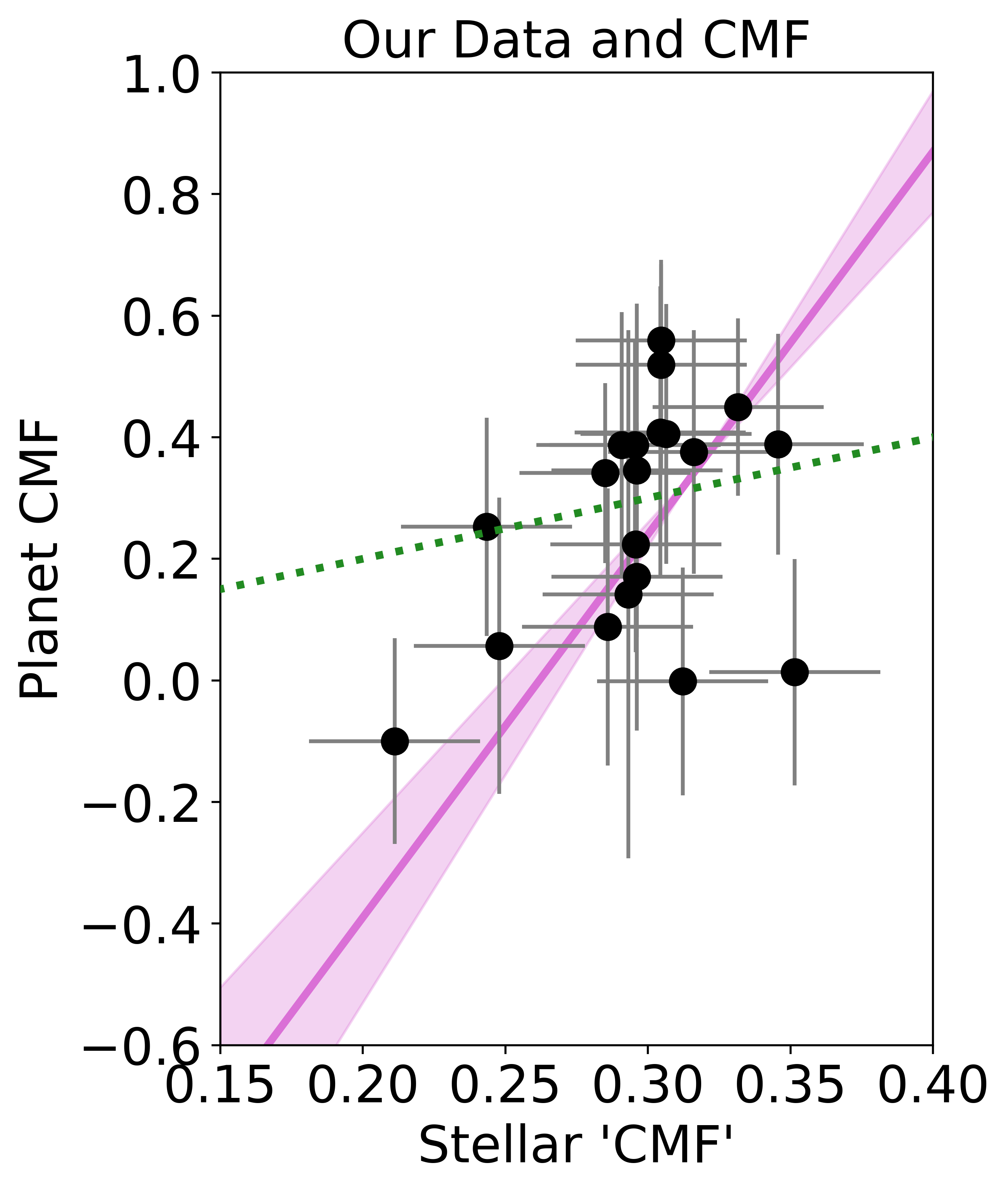}{0.31\textwidth}{}
}
 \caption{Left: Planet CMF and stellar equivalent ``CMF'' for 22 planets (21 host stars) as published in \cite{Adibekyan2021}. Center: Same as left, but using \texttt{SuperEarth} to calculate host star and planet CMF with planet mass and radius values (along with host star abundances) from \cite{Adibekyan2021}. Right: Our \texttt{KeckSpec} sample of 20 planets using planet mass and radius, along with stellar abundances, presented in this work and using \texttt{SuperEarth} to calculate host star and planet CMF. The pink line in each plot is the best-fit slope and 1$\sigma$ uncertainties from \cite{Adibekyan2021}, and the green dotted line is the 1-1 line where stars and planets have the same composition.}
    \label{fig:cmf_comparison}
\end{figure*}

It has been suggested that the overabundance of iron-rich super-Mercury planets around more metal-rich stars could be a product of increased frequency of planet formation, and therefore an increased likelihood of giant impacts \citep{Adibekyan2021}. We note, however, that the planets with a larger CMF than their host star (K2-229 b, Kepler-406 b) orbit host stars of average stellar CMF. 

\subsection{Where are the Super-Mercuries?}

We investigated whether the lack of super-Mercuries in comparison to \cite{Adibekyan2021} is due to planet mass and radius measurements or differing methods of calculating planet and host star CMF. To do this, we took the planet mass and radius measurements for the sample of rocky planets published in \cite{Adibekyan2021} along with their host star abundance measurements and used \texttt{SuperEarth} to calculate the planet and host star equivalent CMF in the same manner described in Sections \ref{sec:CMF} and \ref{sec:Star_CMF}. We compared the values for iron mass fraction in host star (f$^{star}_{iron}\%$) and planet (f$^{planet}_{core}\%$) published in \cite{Adibekyan2021}\footnote{Iron Mass Fraction from \cite{Adibekyan2021} are given as percentages out of 100 instead of fractions, so we divided their published values by 100 to have a direct comparison with our Core Mass Fractions.} to those calculated using our methodology (with their measurements) in Figure \ref{fig:cmf_comparison}. 

We see a much larger spread in CMF using our analysis, including multiple planets having negative iron fractions (indicative of a gaseous envelope as discussed above for TOI-561 b and WASP-47 e). We do not make assumptions about atmosphere thickness for these planets, and therefore do not have all positive iron/rock mass fractions as a result. We also see a wider spread in host star equivalent CMF using our analysis, and we see larger uncertainties in both planet CMF and host star equivalent CMF. 

Despite this, we still do see the five super-Mercuries identified by \citep{Adibekyan2021} and a gap between them and the remaining super-Earths, indicating that while we get differing planet CMF values, we preserve this pattern. Lastly, we confirm that all five planets have a CMF >1$\sigma$ larger than that of their host star. This demonstrates that our lack of super-Mercuries and gap is primarily based on differing planet radius and mass values, not the methodology for computing CMF. The CMF for each of these four Super-Mercuries and their host stars present in our sample, calculated using both our measurements and those from \cite{Adibekyan2021} (from center and right panels of figure \ref{fig:cmf_comparison}), can be found in Table \ref{table:mercuries}. 

The driving force of the differences in planet CMF between our analysis and previous publications comes from differences in planet radius. Changes in radius affect the density of small planets more drastically than changes in mass, and as such small changes in planet radius---which can come from either the R$_{P}$/R$_{*}$ or R$_{*}$ measurement---can drastically change a planet's CMF. This effect is seen most strongly in the variance between literature parameters for K2-229 b: a change in R$_{P}$/R$_{*}$ of 0.0118 $\pm$ 0.0002 \citep{2021PSJ.....2..152A} to 0.0140 $\pm$ 0.0005 \citep{2019ApJ...883...79D} changes the CMF of K2-229 b from 1.09 to 0.55. For this analysis we use R$_{P}$/R$_{*}$=0.014 from \cite{2019ApJ...883...79D} because \cite{2021PSJ.....2..152A} report issues with their transit fit for K2-229 b.Of the 20 planets in our sample, 5 have literature values for R$_{P}$/R$_{*}$ that differ by $>1 \sigma$ (HD 136352 b, HD 15337 b, K2-229 b, Kepler-93 b, and Wasp-74 e). \cite{Adibekyan2021} do not report R$_{P}$/R$_{*}$ in addition to planet radius for their sample, but when we compare planet radius we find that our radius for K2-106 b differs from theirs by $>$1$\sigma$. 

Because literature values of R$_{P}$/R$_{*}$ and therefore planet radius can vary by $>1\sigma$, this suggests an underestimation of our uncertainties on these measurements for many planets. This is one reason we compute CMF uncertanties using a conservative Monte Carlo technique that almost certainly overestimates the uncertainties on CMF. We encourage skepticism when interpreting the CMF of these (and any) rocky planets due to the large dependence on planet radius and potentially underestimated uncertainties planet parameters.

% \begin{figure*}
 %   \centering
 %   \includegraphics[width=1.0\textwidth]{AllRVs_AllTime.pdf}
 %   \includegraphics[width=1.0\textwidth]{Multiplanet_CMF.png}
   
 %   \caption{The Core Mass Fractions for all planets in the Solar System as well as TRAPPIST-1 are shown as a function of incident flux (in units of Earth flux). We see similar scatter in CMF amongst most planets in each system, with Mercury uniquely having a larger CMF. This suggests that the planet formation mechanisms responsible for forming Mercury, either through giant impacts or magnetic interaction, did not produce any such planets around TRAPPIST-1.}
%    \label{fig:trappist}
%\end{figure*}

\section{Conclusion}
\label{sec:conclusion}

For a population of rocky exoplanets with homogeneously measured host star abundances, we calculated the mass and radius of each planet using homogeneously determined stellar parameters. We then calculated the Core Mass Fraction (CMF) for each planet and the equivalent ``Core Mass Fraction'' for each star using their iron and magnesium ratios as determined from homogeneous abundance measurements. Our main conclusions are as follows:
\begin{itemize}
    \item There is a lack of iron-rich high CMF planets orbiting around low metallicity stars, while there is a wide range of planet CMFs around stars of high metallicity and high Fe/Mg. 
    \item Different linear fitting algorithms produce significantly different slopes (m=1.3 $\pm$ 1.0 for OLS vs m=5.6 $\pm$ 1.6 for ODR) for the relationship between planet and host star CMF. This, along with the large scatter in planet CMFs leads us to see no compelling evidence for a strong relationship between planet and host star abundances. Using host star composition as a proxy for planet composition should be treated with caution. 
    \item A quarter of the planets in our sample have a CMF that deviates from that of their host stars by $>$1$\sigma$ (2 planets with larger CMF, 3 with smaller CMF), while three quarters have CMFs that are consistent with their host stars to within 1$\sigma$. 
    \item The larger CMF planets (K2-229 b and Kepler-406 b) orbit stars of average Fe/Mg, while low CMF planets (TOI-561 b, WASP-47 e, and Kepler-78 b) orbit stars that span a wide range of Fe/Mg. 
    \item We do not identify a distinct set of high-density super-Mercuries, and find no gap in CMF between super-Mercuries and super-Earths. We find that previously identified super-Mercuries may have smaller CMFs more consistent with their host star abundances, and this is primarily due to updated mass and radius measurements, not methodology. 
\end{itemize}

One of the limiting factors for this analysis is the small sample size of rocky planets with radius and mass measurements with fractional uncertainties of <30$\%$ . Another limiting factor is the fraction of this sample that has detailed stellar abundance measurements. Future efforts to measure refractory abundances in M-dwarfs may prove to significantly increase the sample size on a timescale much shorter than expanding the rocky planet sample. This will greatly expand our ability to compare the compositions of rocky planets and their host stars. However, our ability to precisely constrain planet CMF will remain a challenge until we can measure planet mass and radius more precisely for small planets.

\facilities{Transiting Exoplanet Survey Satellite (TESS), W. M. Keck Observatory, Gemini Observatory}

\software{SuperEarth \citep{Plotnykov2020} NumPy \citep{harris2020array}, Matplotlib \citep{Hunter:2007}, pandas \citep{mckinney-proc-scipy-2010}, Astropy \citep{astropy:2013, astropy:2018, astropy:2022}, SciPy \citep{2020SciPy-NMeth} }

\section{Acknowledgements}

C.L.B. is supported by the National Science Foundation Graduate Research Fellowship under Grant No. 1842402.

C.L.B., L.M.W. and D.H. acknowledge support from National Aeronautics and Space Administration (80NSSC19K0597) issued through the Astrophysics Data Analysis Program. L.M.W. acknowledges support from the NASA Exoplanet Research Program through grant 80NSSC23K0269.

D.H. also acknowledges support from the Alfred P. Sloan Foundation.

C.L.B. and L.M.W. also acknowledge support from NASA Keck Key Stragetic Mission Support Grant No. 80NSSC19K1475.

L.M.W. thanks Bruce Buffett and the CIDER program, funded under the CSEDI Program (EAR-1903727) and the GeoPRISMS Program (OCE-2025195), for useful discussions.

%\noindent

\bibliography{paper}
\bibliographystyle{aasjournal}

\end{document}